\documentclass[aps,prl,twocolumn,showpacs,groupedaddress]{revtex4}
\usepackage{graphicx}
\usepackage{dcolumn}
\usepackage{amsmath}
\usepackage{amsfonts}
\usepackage{amssymb}
\usepackage{sidecap}

\begin{document}
\title{Hidden itinerant-spin extreme in heavily-overdoped La$_{2-x}$Sr$_{x}$CuO$_4$ revealed by dilute Fe doping: A combined neutron scattering and angle-resolved photoemission study}
\author{Rui-Hua He$^{1, 2, 3, *}$\email[Email: ]{ruihuahe@stanford.edu}, M. Fujita$^{4, *}$, M. Enoki$^5$, M. Hashimoto$^{1, 2, 3}$, S. Iikubo$^{4}$, S.-K. Mo$^{1, 2, 3}$, Hong Yao$^1$, T. Adachi$^5$, Y. Koike$^5$, Z. Hussain$^3$, Z.-X. Shen$^{1, 2}$, K. Yamada$^4$}
\address{
$^{1}$Geballe Laboratory for Advanced Materials, Departments of Physics and Applied Physics, Stanford University, Stanford, California 94305, USA\\
$^{2}$Stanford Institute for Materials and Energy Sciences, SLAC National Accelerator Laboratory, Menlo Park, California 94025, USA\\
$^{3}$Advanced Light Source, Lawrence Berkeley National Laboratory, Berkeley, California 94720, USA \\
$^{4}$Advanced Institute for Materials Research and Institute for Materials Research, Tohoku University, Sendai 980-8577, Japan\\
$^{5}$Department of Applied Physics, Tohoku University,
Sendai 980-8579, Japan\\
$^*$These authors contributed equally to this work. 
}

\begin{abstract}
We demonstrated experimentally a direct way to probe a hidden propensity to the formation of spin density wave (SDW) in a non-magnetic metal with strong Fermi surface nesting. Substituting Fe for a tiny amount of Cu (1\%) induced an incommensurate magnetic order below 20 K in heavily-overdoped La$_{2-x}$Sr$_{x}$CuO$_4$ (LSCO). Elastic neutron scattering suggested that this order cannot be ascribed to the localized spins on Cu or doped Fe. Angle-resolved photoemission spectroscopy (ARPES), combined with numerical calculations, revealed a strong Fermi surface nesting inherent in the pristine LSCO that likely drives this order. The heavily-overdoped Fe-doped LSCO thus represents the first plausible example of the long-sought ``itinerant-spin extreme" of cuprates, where the spins of itinerant doped holes define the magnetic ordering ground state. This finding complements the current picture of cuprate spin physics that highlights the predominant role of localized spins at lower dopings. The demonstrated set of methods could potentially apply to studying hidden density-wave instabilities of other ``nested" materials on the verge of density wave ordering.

\end{abstract}
\maketitle

Whether high-$T_c$ superconductivity (HTSC) is a result of the proximity to a Mott insulator or a Fermi-liquid metal is a perpetual question in modern physics. The fluctuations of spins in both phases have been intensively explored in theory and believed to be essential for understanding the pairing mechanism and cuprate phenomenology \cite{stripe:theory:SteveHowToDetect, stripe:theory:VojtaStripeReview}. Experimentally, the undoped Mott phase has long been established as the strong-coupling extreme dominated by the localized Cu spins. In contrast, the opposing weak-coupling extreme expected to be dominated by the itinerant spins of doped holes at high dopings has so far been elusive. Its successful demonstration is crucial for reaching consensus on the equal importance of itinerant spin physics in HTSC. 

A recent quantum oscillations study has made a progress on this issue by revealing unambiguously a weak-coupling Fermi-liquid like ground state in the heavily-overdoped regime \cite{HTSC:QO_Tl2201}. However, a critical step towards the establishment of the itinerant-spin extreme is yet to be made, which has to demonstrate that such (or a similar) Fermi-liquid like electron system has an inherent tendency towards the SDW formation driven by the nesting of its Fermi surface (FS).

\begin{figure}
\centering
\includegraphics[width=0.48\textwidth]{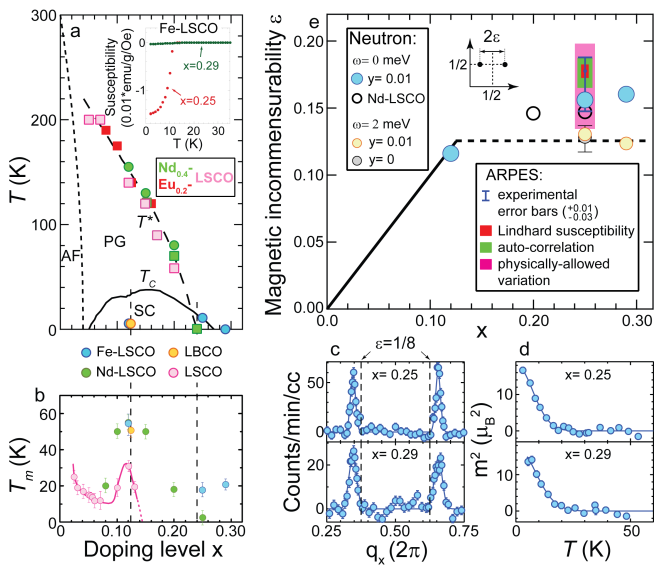}
\caption{(a) Phase diagram of LSCO. The $T^*$ line based on the results of resistivity (circle) and Nernst effect (square) measurements on LSCO, Nd-LSCO \& Eu-doped LSCO (different colors) is reproduced from Ref. \cite{HTSC:Nernst_TailleferRev}. $T_c$ of LSCO (solid line), LBCO and Nd-LSCO (overlapping at x$\sim$ 0.125) are compared with those of Fe-LSCO at x=0.12 \cite{stripe:neutron:FeLSCO1} and 0.25 \& 0.29, determined by magnetic susceptibility measurements (inset). (b) $T_m$ of LSCO \cite{stripe:neutron:LSCOSpinIncomm_diagonalmoment}, Nd-LSCO \cite{stripe:neutron:NdLSCOSpinIncomm}, LBCO \cite{stripe:neutron:LBCO_Fujita}, Fe-LSCO x= 0.12 \cite{stripe:neutron:FeLSCO1}, 0.25 \& 0.29. (c) Elastic magnetic intensity along (-0.25,0.5)-(0.25,0.5) at 10 K after subtracting the background at 60 K. (d) Temperature dependence of $m^2$, evaluated from the resolution-corrected integrated elastic scattering peak intensity. $T_m$ is defined at its 3\% onset. (e) Doping dependence of $\epsilon$. Elastic results of Fe-LSCO and Nd-LSCO \cite{stripe:neutron:NdLSCOSpinIncomm} measured with the energy transfer $\omega=0$ are compared with the inelastic of Fe-LSCO and LSCO \cite{stripe:neutron:LSCOSpinIncomm_vertical} measured with $\omega\sim 2$meV. Error bars (if unseen) are smaller than the symbol size. The color bars at x= 0.25 indicate different incommensurabilities obtained by ARPES and their uncertainties (see Fig. \ref{Fig. 2}g \& i).
}
\label{Fig. 1}
\end{figure}

The reluctance of a magnetic order to appear at a measurable temperature in the heavily-overdoped regime can be naively taken to reflect a weak (if any) nesting. Nevertheless, this is not necessarily true, as a low density wave ordering temperature can be due to the existence of a competing order or the lack of a sizable, proper interaction that is required to mediate the static ordering, even in the case of a strong nesting. Under either circumstance, the itinerant-spin extreme actually exists but remains hidden, unless one can boost the ordering temperature, perform experiments on both the order and electronic states and verify their connection. Such boost can be technically achieved by introducing some perturbations to the system that facilitate the ordering while minimally affect its original nesting.

In this paper, we present the first effort in accomplishing this crucial step towards a complete picture of the cuprate spin physics. We show that the electron system of heavily-overdoped LSCO at x$\geq$ 0.25 likely represents an example of such hidden itinerant-spin extreme. We substituted Fe for only 1\% Cu in these LSCO originally without any magnetic order. We found by elastic neutron scattering that this perturbation was sufficient to induce a robust, quasi-static magnetic order with incommensurate wave vectors below 20 K. This order in La$_{2-x-y}$Sr$_{x+y}$Cu$_{1-y}$Fe$_y$O$_4$ (Fe-LSCO, where x and y= 0.01 are the concentrations of doped holes and Fe) \cite{stripe:neutron:FeLSCO2, stripe:neutron:FeLSCO1} at x$\geq$ 0.25 appeared to be distinct from the well-studied magnetic order in LSCO at lower dopings mainly due to the localized Cu spins. It is also unlikely to be ascribed to the doped Fe spins. Nevertheless, we demonstrated its close link with the itinerant spins of doped holes, by confirming the Fermi-liquid like character of the lowest-lying electronic states with ARPES and revealing a remarkable FS nesting associated with the correct wave vector at x= 0.25. We conclude that the novel magnetic order in Fe-LSCO reflects the hidden instability of the itinerant spins in LSCO at x$\geq$ 0.25 driven by a strong FS nesting, and discuss how the Fe doping promotes their ordering.

\begin{figure*}
\begin{center}
\includegraphics[width=0.78\textwidth]{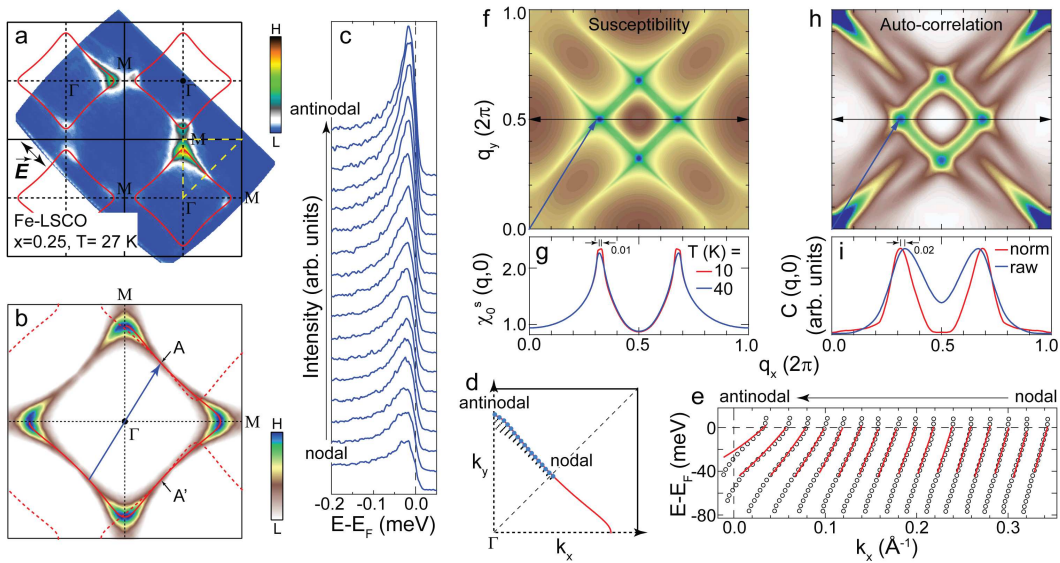}
\end{center}
\vspace*{-0.5cm}
\caption{(a) FS map of Fe-LSCO x= 0.25 over a wide momentum-space region (integrated over $E_F\pm 5$ meV). (b) FS map [$I(\mathbf{k},0)$] symmetrized from the yellow-encircled region in (a) after subtracting a spectral intensity background (30\% of the maximum). Perfect nesting of the FS (solid curve) associated with the blue arrow, $\mathbf{q}_{sh}(0.19)$, is simultaneously achieved around points A \& A'. (c) Spectra at $k_F$ [blue dots in (d)]. (e) Momentum-distribution-curve (MDC) dispersions along black dotted lines (cuts) in (d), subject to the global tight-binding fit (red curves). Note a systematic error in $k_F$ determined by the MDC peak at $E_F$, which corresponds to -0.01 in $\epsilon$ \cite{stripe:other:LBCO_twogap}. (f) Static Lindhard susceptibility, $\chi_0^{s}(\mathbf{q},0)=-\sum_{\mathbf{k}}[f(\varepsilon_{\mathbf{k}+\mathbf{q}}) - f(\varepsilon_{\mathbf{k}})]/[\varepsilon_{\mathbf{k}+\mathbf{q}} - \varepsilon_{\mathbf{k}}]$, where $f()$ is the Fermi function and $\varepsilon_{\mathbf{k}}$ is the band dispersion relation extracted from (e), calculated at $T=20$ K. (g) Susceptibility line profiles along (0,0.5)-(1,0.5) calculated at 10 \& 40 K. (h) The auto-correlation, $C(\mathbf{q},0)\!=\!\sum_{\mathbf{k}}I(\mathbf{k}+\mathbf{q},0)I(\mathbf{k},0)$, of the FS map normalized to the MDC peak intensity at $E_F$. (i) Auto-correlation line profiles along (0,0.5)-(1,0.5) for the raw [(b)] and normalized FS maps. The peak shift specified in (g) [(i)] corresponds to the length of the red (green) bar in Fig. \ref{Fig. 1}e.} 
\label{Fig. 2}
\end{figure*}

Single crystals of Fe-LSCO x= 0.25 \& 0.29 were grown by the travelling-solvent floating-zone method \cite{stripe:neutron:FeLSCO2}, whose $T_c$ ($\pm$1.5 K)=11 K \& 0 K, respectively, were determined by the diamagnetization onset in SQUID magnetometer (Fig. \ref{Fig. 1}a). Neutron scattering experiments were performed on the HER triple-axis spectrometer at the JRR-3M reactor in Tokai, Japan, with the incident neutron energy 5.0 meV, momentum and energy resolutions 0.005 \AA$^{-1}$ and 0.1 meV, respectively. Similar to LSCO, Fe-LSCO x= 0.25 \& 0.29 are in the high-temperature tetragonal phase as indicated by the absence of superlattice peak at $(q_x,q_y,q_z)=(h/2,h/2,l)$ (in unit of $2\pi/a$, $a$= 3.8\AA is the lattice constant; $h$: odd, $l$: even) down to 3 K. ARPES measurements were performed on the ALS beamline 10.0.1 using a Scienta R4000 electron analyzer and 55 eV photons with fixed in-plane polarization \cite{stripe:other:LBCO_twogap}. The angular (momentum) and energy resolutions were 0.25$^\circ$ (0.015 \AA$^{-1}$) and 20 meV. A precise sample alignment (Fig. \ref{Fig. 2}a) yielded an angular uncertainty $\pm 0.1^\circ$.

In LSCO at x$\leq$ 0.125, a quasi-static magnetic order, the so-called spin stripe order \cite{stripe:neutron:TranquadaStripe1st}, has been found by elastic neutron scattering (at energy transfer $\omega= 0$). It has two generic ordering wave vectors, $\mathbf{q}_{sh}(\epsilon)=(0.5\pm\epsilon,0.5)$ ($q_z\equiv 0.5$ hereafter), with the magnetic incommensurability $\epsilon\simeq x$ (the solid line in Fig. \ref{Fig. 1}e) \cite{stripe:neutron:LSCOSpinIncomm_vertical}. Such unusual doping dependence suggests a continuous evolution of the stripe order out of the undoped antiferromagnetic order (with $\epsilon=0$) upon hole doping, likely by forming microscopically segregated, incompressible spin and hole stripes. It has been reproduced by various model calculations in the strong-coupling limit and generally taken to support a localized-spin origin of the
stripe order \cite{stripe:theory:SteveHowToDetect, stripe:theory:VojtaStripeReview}.

Introducing additional impurities into LSCO at x$\leq$ 0.125 commonly stabilizes the stripe order (maintaining the same $\epsilon\simeq x$ scaling), and concomitantly depresses the coexisting superconductivity. This can be seen, for example, with 1\% Fe doping at x= 0.12. Compared with LSCO, Fe-LSCO shows the same $\epsilon$ (Fig. \ref{Fig. 1}e), a much increased magnetic ordering temperature $T_m$ (Fig. \ref{Fig. 1}b) and a strongly suppressed $T_c$ (Fig. \ref{Fig. 1}a), all in quantitative agreement with the results of Nd-doped LSCO (Nd-LSCO) and La$_{2-x}$Ba$_{x}$CuO$_4$ (LBCO) at x$\sim$ 0.125.

The situation for x$>$ 0.125 appears quite different, both in the pristine and impurity-doped LSCO. In LSCO, no elastic scattering intensity was consistently found at the lowest measuring temperature (yielding $T_m\sim 0$, Fig. \ref{Fig. 1}b). The $\epsilon$ defined by the finite \textit{inelastic} scattering signal at $\omega\sim 2$ meV appears doping independent and close to 0.125. In Nd-LSCO, elastic scattering intensity also decreases above x= 0.12 but remains substantial up to x= 0.20, until it becomes barely detectable at x= 0.25 (note $T_m$ in Fig. \ref{Fig. 1}b) \cite{stripe:neutron:NdLSCOSpinIncomm}. The $\epsilon$ defined \textit{elastically} deviates considerably from both 0.125 and the $\epsilon\simeq x$ scaling (Fig. \ref{Fig. 1}e). Such deviation, if proven to be intrinsic, could suggest different spin physics at play in the ground state of LSCO at x$>$ 0.125. Nevertheless, this issue is complicated by the high Nd impurity concentration (20\%), modification of the crystal structure at low temperatures and smallness of the observed elastic signal at x= 0.25.

These concerns do not apply to Fe-LSCO x= 0.25 \& 0.29, in which a robust quasi-static magnetic order is induced by only 1\% Fe doping without changing the crystal structure. The temperature dependence of the ordered magnetic moment $m$ shows a similar onset at $T_m\sim 20$ K and a comparable low-temperature maximum for both dopings (Fig. \ref{Fig. 1}d). The persistence of this order in the heavily-overdoped regime contrasts the weakening of the magnetic order as doping increases above 0.125 in LSCO and Nd-LSCO (Fig. \ref{Fig. 1}b). Its inducement also appears to have a much weaker effect on the coexisting superconductivity (Fig. \ref{Fig. 1}a). Remarkably, such robust magnetic order is associated with an $\epsilon$ that is clearly distinct from both 0.125 and the $\epsilon\simeq x$ scaling, but consistent with that of Nd-LSCO at x= 0.25 (Fig. \ref{Fig. 1}e). Meanwhile, inelastic scattering measurements of these samples found similar low-energy magnetic fluctuations as in LSCO, as exemplified by $\epsilon\sim$ 0.125 at $\omega\sim 2$ meV in Fig. \ref{Fig. 1}e.

This magnetic order cannot be simply ascribed to the doped Fe spins. The ordered magnetic moment per unit cell is $m\sim$0.13 $\mu_B$. The maximal possible contribution from the doped Fe is 0.05 $\mu_B$ (=1\% of 5 $\mu_B$), which assumes the same periodic arrangement of the Fe spins as the magnetic order with a period $\sim 6.4a$. A more realistic situation is that these spins localized on the Fe sites are distributed randomly or evenly but with a different period $\bar{d}_{Fe-Fe}\sim 10a$. Either case suggests a negligible effective contribution of the Fe spins to $m$. Moreover, our Fe-doping-dependent study suggests that the $\epsilon$ barely changes with $y$ increasing from 0.01 to 0.08, which is inconsistent with a responsible ordering of pure Fe spins. 

Taken collectively, the above results suggest that the novel Fe-induced magnetic ordering is an intrinsic yet latent ground state property of heavily-overdoped LSCO, which appears different from the (stabilized) stripe ordering at lower dopings. While neutron scattering alone does not rule out its possible connection with a different form of ordering of the localized Cu spins, complementary ARPES results suggest that it more likely originates from the spins of itinerant doped holes. 

ARPES measurement on Fe-LSCO x= 0.25 above $T_m$ revealed well-defined quasiparticles along the entire FS associated with the doped holes (Fig. \ref{Fig. 2}c). While this is reminiscent of the generalized Fermi-liquid extreme established in heavily-overdoped Tl$_2$Ba$_2$CuO$_{6+\delta}$ \cite{HTSC:QO_Tl2201}, it appears very different from x= 0.12, where a large pseudogap exists in the antinodal region making the quasiparticles therein ill-defined \cite{stripe:other:LBCO_twogap}. In LSCO, the pseudogap formation temperature ($T^*$) decreases with overdoping and goes to zero at x$\sim$ 0.24, disregarding the type and amount of dopants (Fig. \ref{Fig. 1}a) \cite{HTSC:Nernst_TailleferRev}. Recent progresses indicated that finite dynamic stripe correlations start to develop below $T^*$ \cite{HTSC:Nernst_TailleferRev, stripe:neutron:YBCO_eLiquidCrystal, HTSC:PseudogapDispersion, stripe:STM:PseudogapFluctStripe_Bi2212}. Our ARPES data are consistent with the absence of the pseudogap and stripe correlations in (Fe-) LSCO at x$\geq$ 0.25, and support the dominance of itinerant physics in the ground state.

We further examine the tendency of the itinerant quasiparticles towards density wave formation. We performed both the model-independent ARPES auto-correlation \cite{CDW:DaweiNaTaS} (Fig. \ref{Fig. 2}h \& i) of the experimental FS (Fig. \ref{Fig. 2}b) and the calculation of Lindhard susceptibility  \cite{CDW:SDW_Cr_RPA, CDW:HongSusceptibility, CDW:BorisenkoFSnesting} (Fig. \ref{Fig. 2}f \& g) based on the experimental quasiparticle band structure (Fig. \ref{Fig. 2}d \& e). 

The ARPES auto-correlation shows four local maxima at $\mathbf{q}_{sh}(\epsilon_{AC})$ and $\mathbf{q}_{sv}(\epsilon_{AC})=(0.5,0.5\pm\epsilon_{AC})$. The locations of these auto-correlation peaks are mainly determined by the FS shape, but affected by the spectral weight distribution or quasiparticle broadening along and normal to the FS. The exact value of $\epsilon_{AC}$ varies slightly among different schemes for the normalization and/or background subtraction of the spectral intensity, with a typical variation specified in Fig. \ref{Fig. 2}i. 

Lindhard susceptibility shows four global maxima at $\mathbf{q}_{sh}(\epsilon_{sus})$ and $\mathbf{q}_{sv}(\epsilon_{sus})$. These susceptibility peaks are given by the nesting of the FS, which can be spanned by the corresponding wave vectors (e.g., the blue arrow in Fig. \ref{Fig. 2}b). They only slightly shifted by the thermal smearing, as exemplified in Fig. \ref{Fig. 2}g. Note that the susceptibility peak features are much sharper than those by the ARPES auto-correlation. This is because the Lindhard function is derived for non-interacting (bare) electrons unlike the actual interacting system with a finite quasiparticle lifetime. Such singular $\mathbf{q}$-space landscape of the bare spin susceptibility is highly non-trivial and indeed remarkable, even compared with those of the well-known weak-coupling density wave prototypes, such as Cr \cite{CDW:SDW_Cr_RPA}, rare-earth tritellurides \cite{CDW:HongSusceptibility} and transition-metal dichalcogenides \cite{CDW:BorisenkoFSnesting}. Inherent of such a strong FS nesting, the itinerant quasiparticle system is expected to develop a (spin or charge) density wave order with the ordering wave vectors $\mathbf{q}_{sh}$ and/or $\mathbf{q}_{sv}$ at a finite temperature, given the existence of a finite residual (repulsive or attractive) interaction between quasiparticles. 

As summarized in Fig. \ref{Fig. 1}e, both $\epsilon_{AC}$ and $\epsilon_{sus}$ determined by ARPES on Fe-LSCO x= 0.25 match the $\epsilon$ measured by neutrons reasonably well, with the ARPES experimental error bars and physical variations due to finite thermal smearing and/or quasiparticle broadening taken into account. Combining this with the neutron results, we conclude that the novel magnetic order in heavily-overdoped Fe-LSCO likely arises from the itinerant spins.

It is reasonable to believe that the dilute Fe doping does not severely affect the original nesting property of LSCO. Our ARPES measurement on the pristine LSCO showed the same FS within experimental uncertainty as the Fe-doped one and a FS volume consistent with a previous report \cite{cuprates:LSCO:TeppeiReview}. Therefore, the physical character and band structure of the itinerant quasiparticles in LSCO are unlikely changed. Impurity scattering is generally detrimental to any nesting instability by contributing additional smearing to the susceptibility peak. But this might not be worrisome for our case with a low Fe doping level and singular susceptibility peaks. 

Albeit small, such perturbation turns out to be effective in enforcing the ordering of the itinerant spins. This can in principle be achieved by suppressing the competing superconductivity or, alternatively, strengthening or introducing their repulsive interaction of some form. The first possibility can be ruled out in our case as the Fe doping does not change $T_c$ appreciably at x= 0.25 and no superconductivity is originally present at x= 0.29. 

About the second possibility, we first note that impurities in general can cause the slowing down of existing stripe or density wave fluctuations in their vicinities, and lead to their associated static orders depending on the strength of electron correlations \cite{stripe:theory:DisordereffectStripeSDW}. Such general argument is consistent with the observations of impurity-induced static magnetic orders in various cuprate families (to cite a few, Refs. \cite{stripe:neutron:FeLSCO1, stripe:neutron:FeBi2201_1, stripe:neutron:mSR_LSCO_Zn, stripe:neutron:YBCO_Zn}). It can also rationalize why the same level of Fe doping in LSCO stabilizes the stripe order at x= 0.12 but induces a SDW at x$\geq$ 0.25. However, it does not explain why the stabilization/inducement of magnetic orders in LSCO appears overall more effective with the magnetic impurities (e.g., Nd and Fe) than the non-magnetic (e.g., Zn, Gd) \cite{stripe:neutron:FeLSCO1}. In this context, we note that a RKKY-type mechanism specific to magnetic impurities \cite{CDW:SDW_RKKY_CuMn} has recently been invoked to explain the inducement of a static magnetic order in Bi2201 by heavier Fe dopings \cite{stripe:neutron:FeBi2201_1}. The magnetic field effects on both the order and resistivity were there found to be distinct from those of the stripe order \cite{stripe:neutron:FeBi2201_2}. The possible commonality of the Fe-induced magnetic ordering in both cuprate systems, in terms of its microscopic mechanism, the itinerant-spin nature and field dependence, is an important subject for further investigations.

Our finding provides so far the most promising experimental basis for the weak-coupling itinerant-spin extreme of cuprates that has been extensively assumed in and underlaid continued theoretical discussions (e.g, Refs. \cite{stripe:theory:BCSsusceptibility, stripe:theory:Susceptibility_Levy1, stripe:theory:Susceptibility_Littlewood, stripe:theory:BCSsusceptibility_Norman}, see more cited by Refs. \cite{stripe:theory:SteveHowToDetect, stripe:theory:VojtaStripeReview}). Such weak-coupling aspect of the ground-state spin physics, if confirmed to be general for heavily-overdoped cuprates, likely persists and mixes up with its strong-coupling counterpart in the intermediate ranges of doping and energy (transfer), hence being implicated with the rich, yet still puzzling cuprate phenomenology observed therein.

The authors would like to acknowledge inspiring discussions with L. Taillefer and S. A. Kivelson. The work at Tohoku is supported by Grant-In-Aid for Scientific Research (A) (22244039) and (C) (20540342) from the MEXT (Japan). The work at Stanford and ALS is supported by the DOE Office of BES under contracts DE-AC02-76SF00515 and DE-AC02-05CH11231.


\begin{thebibliography}{34}

\bibitem{stripe:theory:SteveHowToDetect}
S. A. Kivelson {\sl et al.} \emph{Rev. Mod. Phys.}, \textbf{75,} 1201 (2003).


\bibitem{stripe:theory:VojtaStripeReview}
M. Vojta, \emph{Adv. Phys.} \textbf{58,} 699 (2009).

\bibitem{HTSC:QO_Tl2201}
B. Vignolle {\sl et al.}, \emph{Nature} \textbf{455,} 952 (2008).

\bibitem{stripe:neutron:FeLSCO2}
M. Fujita {\sl et al.}, \emph{J. Phys. Chem. Solid} \textbf{69,} 3167 (2008).

\bibitem{stripe:neutron:FeLSCO1}
M. Fujita {\sl et al.}, arXiv:0903.5391 (unpublished).

\bibitem{stripe:other:LBCO_twogap}
R.-H. He  {\sl et al.}, \emph{Nature Phys.} \textbf{5,} 119 (2009).

\bibitem{stripe:neutron:TranquadaStripe1st}
J. M. Tranquada {\sl et al.}, \emph{Nature} \textbf{375,} 561 (1995).

\bibitem{stripe:neutron:LSCOSpinIncomm_vertical}
K. Yamada {\sl et al.}, \emph{Phys. Rev. B} \textbf{57,} 6165 (1998).


\bibitem{stripe:neutron:NdLSCOSpinIncomm}
N. Ichikawa {\sl et al.}, \emph{Phys. Rev. Lett.} \textbf{85,} 1738 (2000).

\bibitem{HTSC:Nernst_TailleferRev}
L. Taillefer, \emph{Annu. Rev. Condens. Matter Phys.} \textbf{1,} 10 (2010).

\bibitem{stripe:neutron:YBCO_eLiquidCrystal}
V. Hinkov {\sl et al.}, \emph{Science} \textbf{319,} 597 (2008).



\bibitem{HTSC:PseudogapDispersion}
M. Hashimoto {\sl et al.}, \emph{Nature Phys.} \textbf{6,} 414 (2010).

\bibitem{stripe:STM:PseudogapFluctStripe_Bi2212}
C. V. Parker {\sl et al.}, \emph{Nature} \textbf{468,} 677 (2010).


\bibitem{CDW:DaweiNaTaS}
D. W. Shen {\sl et al.}, \emph{Phys. Rev. Lett.} \textbf{99,} 216404 (2007).

\bibitem{CDW:SDW_Cr_RPA}
K. Schwartzman, J. L. Fry \& Y. Z. Zhao, \emph{Phys. Rev. B} \textbf{40,} 454 (1989).

\bibitem{CDW:HongSusceptibility}
H. Yao {\sl et al.}, \emph{Phys. Rev. B} \textbf{74,} 245126 (2006).

\bibitem{CDW:BorisenkoFSnesting}
D. S. Inosov {\sl et al.}, \emph{New J. Phys.} \textbf{10,} 125027 (2008).

\bibitem{cuprates:LSCO:TeppeiReview}
T. Yoshida {\sl et al.}, \emph{J. Phys.: Condens. Matter} \textbf{19,} 125209 (2007).

\bibitem{stripe:theory:DisordereffectStripeSDW}
B. M. Andersen, S. Graser \& P. J. Hirschfeld, \emph{Phys. Rev. Lett.} \textbf{105,} 147002 (2010).

\bibitem{stripe:neutron:FeBi2201_1}
H. Hiraka {\sl et al.}, \emph{Phys. Rev. B} \textbf{81,} 144501 (2010).

\bibitem{stripe:neutron:mSR_LSCO_Zn}
C. Panagopoulos {\sl et al.}, \emph{Phys. Rev. B} \textbf{69,} 144510 (2004).

\bibitem{stripe:neutron:YBCO_Zn}
A. Suchaneck {\sl et al.}, \emph{Phys. Rev. Lett.} \textbf{105,} 037207 (2010).

\bibitem{CDW:SDW_RKKY_CuMn}
F. J. Lamelas {\sl et al.}, \emph{Phys. Rev. B} \textbf{51,} 621 (1995).

\bibitem{stripe:neutron:FeBi2201_2}
S. Wakimoto {\sl et al.}, \emph{Phys. Rev. B} \textbf{82,} 064507 (2010).

\bibitem{stripe:theory:BCSsusceptibility}
N. Bulut, D. J. Scalapino, \emph{Phys. Rev. B} \textbf{45,} 2371 (1992).

\bibitem{stripe:theory:Susceptibility_Levy1}
Q. Si {\sl et al.}, \emph{Phys. Rev. B} \textbf{47,} 9055 (1993).

\bibitem{stripe:theory:Susceptibility_Littlewood}
P. B. Littlewood {\sl et al.}, \emph{Phys. Rev. B} \textbf{48,} 487 (1993).

\bibitem{stripe:theory:BCSsusceptibility_Norman}
M. R. Norman, \emph{Phys. Rev. B} \textbf{61,} 14751 (2000).



\bibitem{stripe:neutron:LSCOSpinIncomm_diagonalmoment}
S. Wakimoto {\sl et al.}, \emph{Phys. Rev. B} \textbf{63,} 172501 (2001).

\bibitem{stripe:neutron:LBCO_Fujita}
M. Fujita {\sl et al.} \emph{Phys. Rev. B} \textbf{70}, 104517 (2004).

\end{thebibliography}
\end{document}